\documentclass[aps,prl,twocolumn,groupedaddress,showpacs,amsmath,amssymb]{revtex4}
\usepackage{graphicx}
\usepackage{bm}

\begin{document}
\preprint{prl/vortex-VWR}

\title{Turbulence in Boundary Flow of Superfluid $^4$He Triggered by
Free Vortex Rings}

\author{R. Goto} \author{S. Fujiyama} \author{H. Yano}
\email{hideo@sci.osaka-cu.ac.jp} \author{Y. Nago} \author{N. Hashimoto}
\author{K. Obara} \author{O. Ishikawa} \author{M. Tsubota}
\author{T. Hata} \affiliation{Graduate School of Science, Osaka City
University, Osaka 558-8585, Japan}

\date{\today}

\begin{abstract}
 The transition to turbulence in the boundary flow of superfluid $^4$He
 is investigated using a vortex--free vibrating wire. At high wire
 vibration velocities, we found that stable alternating flow around the
 wire enters a turbulent phase triggered by free vortex rings. Numerical
 simulations of vortex dynamics demonstrate that vortex rings can attach
 to the surface of an oscillating obstacle and expand unstably due to
 the boundary flow of the superfluid, forming turbulence. Experimental
 investigations indicate that the turbulent phase continues even after
 stopping the injection of vortex rings, which is also confirmed by the
 simulations.
\end{abstract}

\maketitle


Quantum turbulence in superfluid $^4$He consists of quantized vortices
that have a simple structure: a normal fluid core and a quantized
circulation \cite{Don-b}. Since this structure prevents the core from
becoming a superfluid, vortices in a superfluid adopt the form of a ring
or a line attached to boundaries. A vortex ring propagates in a
superfluid at a self-induced velocity and eventually attaches to a
boundary. If boundary flow is applied, it is likely that it will cause a
vortex line attached to a boundary to expand \cite{GD}, forming
turbulence. However, this simple transition to turbulence triggered by
vortex rings has not been observed experimentally since remanent
vortices always remain attached to boundaries.

Quantized vortices may nucleate during transition to the superfluid
state \cite{Be2006}, and counterflow occurring during cooling can force
vortices to expand, causing the number of vortices to multiply
\cite{Don-b}. Free vortices that are not attached to boundaries
dissipate by mutual friction or by reconnection of vortices
\cite{Schw1988,Kob2005}, while vortices remain attached between
boundaries \cite{Schw1984}. Therefore, an oscillating obstacle such as a
sphere \cite{Sch1995}, grid \cite{Nic-prl}, or wire
\cite{Ya2006lt,Ya2007prb} can always generate turbulence in superfluid
$^4$He around an oscillating velocity of 50 mm/s, much lower than the
Landau velocity of about 60 m/s at which the superfluidity collapses
\cite{Don-b}. It appears to be difficult to establish a vortex-free
state on the surface of an obstacle in superfluid $^4$He, in which the
critical velocity for the transition to turbulence is expected to be
very high. A vortex--free state has only been achieved in the vicinity
of a moving ion, with vortex nucleation occurring at velocities
exceeding 10 m/s \cite{Mey1961}.

In a recent study \cite{Ya2007prb2}, we reported that slow filling of a
superfluid and using a filter of vortices can prevent vortices from
attaching to boundaries, which prevents the generation of turbulence by
oscillation, even above a velocity of 1 m/s. This method enables the
formation of an oscillating obstacle whose surface is effectively free
of remanent vortices. In the present paper, we report experimental and
numerical investigations of the responses of a vortex--free oscillating
obstacle sprinkled with vortex rings. In the experiments, we
successfully observed a transition to turbulence triggered by vortex
rings using a vortex--free vibrating wire. Numerical simulations based
on a vortex filament model revealed a new process, namely, the
turbulence is generated by vortex rings propagating to the oscillating
obstacle, where the oscillations causes them to grow and then sustain
them.

\begin{figure}[b]
\includegraphics[width=0.6\linewidth]{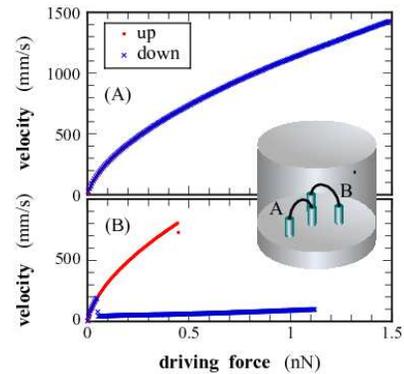} \caption{\label{vwrs}(color
online) Peak velocities of two vibrating wires in superfluid $^4$He
measured at 30 mK for up and down sweeps of the driving force: (A) no
transition to turbulence; (B) generation of turbulence. The inset shows
two vibrating wires (A and B) in a chamber with a pinhole. Graphs (A)
and (B) were obtained using wires A and B, respectively. Wire A was used
as the detector, and wire B as the generator (see text).}
\end{figure}

The experimental setup consisted of two vibrating wires, similar to
those described in a previous paper \cite{Ya2007prb2}. These 3-$\mu$m
NbTi superconducting wires, were formed into semicircles, the two arms
of which were attached to columns mounted on a copper plate, as shown in
the inset in Fig.~\ref{vwrs}.
The wires labeled A and B in Fig.~\ref{vwrs} were located in a small
chamber with a 0.1-mm-diameter pinhole. Wire A was oriented so that it
pointed towards the center of wire B, perpendicular to its semicircle
face. The distance between the arms was 1.6 mm for wire A and 1.8 mm for
wire B. The heights of the columns could be varied so as to adjust the
heights of the wires. The distance between the apexes of the wire loops
was 1.8 mm. The wires and the chamber were located in a cell in a
magnetic field of 25 mT. A heat exchanger made of sintered silver powder
was mounted in the bottom of the cell for cooling the helium. A RuO$_2$
thermometer was mounted on the cell wall. The responses of the vibrating
wires were measured in vacuum at 30 mK. The resonance frequencies were
1590 Hz for wire A and 1030 Hz for wire B. In the present paper, we
report experimental results obtained at 30 mK.


In the experiment, the cell was filled with superfluid $^4$He cooled
below 100 mK over 48 h. This careful filling prevented the formation of
remanent vortices. The pinhole of the chamber also had the effect of
filtering vortices \cite{Ya2007prb2}. As Fig.~\ref{vwrs}(A) shows,
vibrating wire A was unable to generate turbulence in the superfluid. To
test the stability of the superfluid, we applied mechanical vibrations
to the cryostat and warmed the superfluid to 1 K. However, even under
these conditions the vibration of the wire was too stable to collapse,
indicating that no turbulence was generated. In contrast, vibrating wire
B generated turbulence, as Fig.~\ref{vwrs}(B) shows. It appears that
vortices remain attached to wire B. Thus, we were able to produce both a
vortex--free wire and a vortex--attached wire in a chamber.

Turbulence in a superfluid generated by an oscillating obstacle is
accompanied by free vortex rings \cite{Fi2001,Fi2005}. Therefore, many
vortex rings are expected to be generated in the chamber by the
vibration of wire B, and these vortex rings can propagate through the
surrounding fluid and attach to the surface of the wires. However, after
the vibration of wire B was stopped, we found that the velocity of wire
A exhibited the same dependence as that shown in Fig.~\ref{vwrs}(A),
indicating that there was no transition to turbulence. The vortices
generated by wire B seem to be unable to form remanent vortices
attached to wire A. A previous study \cite{Ya2007prb2} reported that
vortices that nucleate during a superfluid transition form bridges
between an obstacle and the surrounding wall to become remanent
vortices. If these bridge vortices are forced to oscillate largely by
the vibration of the obstacle, free vortex rings can be generated by
reconnection of the vortices, forming turbulence \cite{Han2007}. Free
vortex rings in turbulence are expected to be too small to form bridges
between a wire and the surrounding wall. Vortices attaching to the
surface of wire A might vanish if there is no boundary flow. This
condition is suitable for studying the transition to turbulence
triggered by free vortex rings. Wire A can be used to detect the
transition to turbulence, and wire B can be used to generate free vortex
rings.

\begin{figure}
\includegraphics[width=0.65\linewidth]{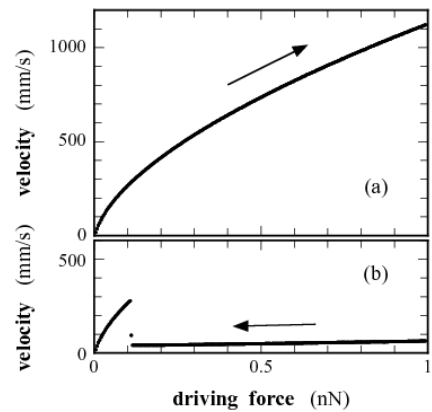} \caption{\label{transition}
Velocities of the detector (wire A) (a) before and (b) after generation
of turbulence by the generator (wire B). The data in (a) were obtained
using an up sweep of the driving force. The turbulence was triggered at
a driving force of 1 nN. After a transition to turbulence, data was
acquired again with a down sweep, as shown in (b) (see text).}
\end{figure}

After stopping the vibration of the generator (wire B), we increased the
velocity of the detector (wire A) above 1 m/s, as shown in
Fig.~\ref{transition}(a).
Maintaining the driving force of the detector at 1 nN, we increased the
velocity of the generator so as to generate turbulence. At the moment of
transition to turbulence, the detector's velocity suddenly dropped to a
low value of around 60 mm/s. This result indicates that free vortex
rings were generated by the generator, propagating to the detector and
interfering with its vibration. We again stopped the vibration of the
generator, but the detector's velocity still remained low, indicating
that the detector had entered a turbulent phase. This is a surprising
result because free vortex rings not only disturb the wire vibration but
also trigger a turbulent phase. During the transition to turbulence,
free vortex rings may attach to the surface of the detector, expanding
due to boundary flow and forming turbulence. Even after the generator
ceases to generate free vortex rings, vortices attached to the
detector's surface may still develop by the flow and form
turbulence. As the driving force was reduced, the velocity of the detector
remained low until the flow around it entered a laminar phase, as shown
in Fig.~\ref{transition}(b). While the generator was not vibrating, we
further increased the driving force of the detector and found that the
detector's velocity traces the initial values shown in
Fig.~\ref{transition}(a), indicating that there was no turbulence. Vortices
nucleating in the turbulent phase were unable to affect the condition of
the detector.

It has been reported that free vortex rings can affect the vibration of
a wire in superfluids \cite{Fi2001,Fi2005jltp}. In superfluid $^4$He,
free vortex rings disturb the vibration of a wire, decreasing the
critical velocity of turbulence \cite{Fi2005jltp}. This result implies
that vortex rings might trigger a transition to turbulence. However, the
wire could generate turbulence independently, indicating that remanent
vortices were attached to the wire from the beginning. Under this
condition, it is difficult to understand how vortex rings cause
turbulence, since the remanent vortices would also affect the transition
to turbulence. In the present study, however, the absence of remanent
vortices clarifies this difficulty. In $^3$He--B, instead of disturbing
the vibration of a wire, vortex rings rather increase it \cite{Fi2001},
since vortices can shield the wire from thermal excitations in the
superfluid. This shielding effect may limit observations of dissipation.

\begin{figure}
 \includegraphics[width=0.78\linewidth]{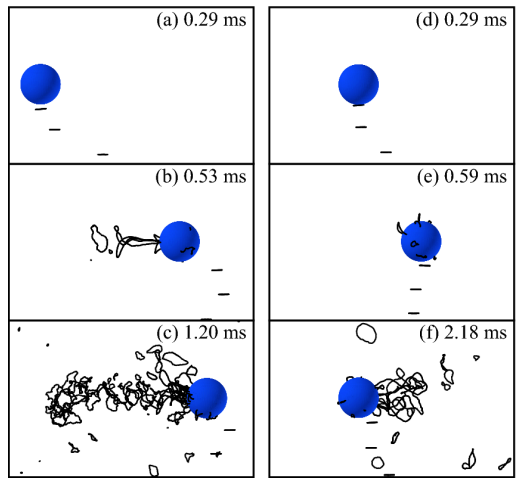} \caption{(color online)
 Snapshots of the time evolution of vortex tangle from the start of
 vortex-ring injection: (a), (b), and (c) are for an oscillation
 velocity of 137 mm/s, and (d), (e), and (f) are for 50 mm/s. The blue
 solid circle is an oscillating sphere. In (a) and (d), the three lines
 below the sphere are vortex rings injected from the bottom.}
 \label{snapshot}
\end{figure}

These observations were verified by performing numerical simulations. In
modeling the experimental situation, an oscillating sphere was
considered rather than a vibrating wire; this is because it is easier to
calculate the case of a sphere, while it still reflects the essence of
the physics that we are interested in.  The problems we desire to solve
numerically are what occurs when vortex rings coming from the generator
collide with an oscillating sphere, and whether after collision these
vortices form turbulence or not. The vortex filament model is used in
the simulation \cite{Schwarz,Tsubota}. The superfluid velocity
$\bm{v_{vor}}$ generated by the vortices is calculated using Biot-Savart
integration. An additional velocity field $\bm{v_b}$ is imposed in order
to satisfy the boundary condition on the surface of the sphere for a
superfluid $(\bm{v_s}-\bm{v_p})\cdot \bm{n}=0$, where
$\bm{v_s}=\bm{v_{vor}}+\bm{v_b}$ is the total superfluid velocity,
$\bm{v_p}$ is the sphere velocity, and $\bm{n}$ is the unit normal
vector to the surface of the sphere. The additional velocity $\bm{v_b}$
takes the form
\begin{equation}
\bm{v_b}= \nabla\Phi _b + \nabla\Phi _u,
\end{equation}
where $\Phi_b$ is the velocity potential that cancels the normal
velocity component to the sphere made by the vortices \cite{Schwarz} and
\begin{equation}
\Phi _u = -\frac{1}{2} \left(\frac{R}{r}\right)^3 \bm{v_p}\cdot\bm{r} \label{sph_field}
\end{equation}
is the scalar velocity potential made by the moving sphere. Here $R$
denotes the radius of the sphere, and $\bm{r}$ is the position from the
center of the sphere. The reconnection process is assumed to occur when
a vortex becomes close to another vortex or a spherical boundary within
the computational resolution. The parameters used in the calculation
were chosen to match those of the experiments, such as a sphere radius
of 3 $\mu$m and an oscillation frequency of 1590 Hz. The velocity of the
oscillation is chosen to be 137 mm/s which is much higher than the
observed critical velocity, 50 mm/s close to the velocity, and 30 mm/s
much lower than the velocity. The experiments were performed at a very
low temperature of 30 mK, where the normal fluid component is
negligible. Thus, we performed the numerical simulation at the zero
temperature limit to neglect mutual friction. Vortex rings all having a
radius of 0.74 $\mu$m \cite{ring_size} were injected every 0.05 ms
vertically from the bottom \cite{ring_frequency}
(Fig.~\ref{snapshot}(a)).

The numerical simulation reveals a new explanation for the transition
from vortex rings to turbulence. The incoming vortex rings collide
successively with the sphere. The vortices are then split into lots of
segments of vortex loops that attach to the sphere. Smaller vortex loops
than the sphere radius will be split into smaller loops by reconnection
during running on its surface, eventually vanishing. In contrast, large
loops still remain attached to the sphere. The ends of a large vortex
loop are dragged by a sphere motion; however, the other parts of the
loop can move with each self-induced velocity rather independently.
Consequently, the sphere motion stretches a vortex line, up to a length
of the oscillation amplitude (Fig.~\ref{snapshot}(b)). Some of the large
vortices that have grown sufficiently detach from the sphere. Since they
have a low self-induced velocity, they are unable to escape before the
sphere returns. These vortices are struck by the sphere many times, so
that vortex segments become attached to the sphere. These processes are
repeated with the sphere oscillation. The remaining vortex segments
start to grow, and finally the sphere becomes covered with a vortex
tangle (Fig.~\ref{snapshot}(c)). The volume that the created vortex
tangle occupies is smaller in the simulation for an oscillation velocity
of 50 mm/s than for that in the case of 137 mm/s; this is due to the
different oscillation amplitudes of these two simulations, since the
sphere can stretch vortices only up to the oscillation amplitude
(Figs.~\ref{snapshot}(c) and (f)). For 30 mm/s, in contrast, though the
vortices that had collided with the sphere grew slightly by the
oscillation, they reconnect immediately to themselves, flying away from
the sphere. Therefore, vortices will disappear soon after stopping the
injection of vortex rings. The roughness of the sphere surface should be
concerned with respect to the stretch of a vortex loop. The roughness
could affect only the motion of loop ends attached to the surface, thus
not changing much the present scenario.

As mentioned in the experimental section, the vortex tangle in the
detector survives even after the vortex ring supply from the generator
has been stopped. To confirm this behavior numerically, the vortex rings
supply was switched off after eight vortices had been injected into the
fluid. The time development for this case is very similar to that for
the previous case. Collisions between large vortices and the sphere
leave many vortex segments attached to the sphere's surface. These
segments behave as seeds for turbulence, and start to grow. As a result,
the vortex tangle is sustained by the sphere even after the supply of
vortex rings has stopped, which is consistent with the observations.

 \begin{figure}
 \includegraphics[width=0.8\linewidth]{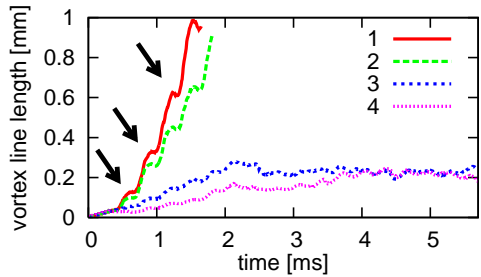} \caption{(color
 online) Time evolution of vortex lines for an oscillating velocity of
 137 mm/s with continuous vortex--rings injection (red solid line, 1),
 for 137 mm/s with the injection of eight vortex--rings (green dashed
 line, 2), for 50 mm/s with continuous vortex--ring injection (blue
 dotted line, 3), and for 50 mm/s with the injection of eight
 vortex--rings (pink dotted line, 4). The arrows indicate cusps on the
 solid line (see text).}  \label{vorlen}
\end{figure}
Figure \ref{vorlen} shows the time development of the vortex
length. Since we are interested in only the volume around the sphere
covered with a tangle, the vortices that exit the computational box (25
$\mu$m)$^3$ are not considered when evaluating the vortex length. The
increase in the vortex length until 0.8 ms is due to the injection of
vortex rings. The cusps of lines observed at every 0.3 ms in
Fig.~\ref{vorlen} indicate that vortex growth temporarily ceases when
the sphere loses its velocity in oscillation. It is clearly confirmed
that for both 50 mm/s and 137 mm/s, the vortex tangle is sustained even
when the vortex supply from the generator ceases. In the case 50 mm/s,
the vortex growth produced by the oscillation is balanced by the vortex
loss due to escape, which causes the vortex line length to be saturated
after 2 ms. For a velocity of 137 mm/s, technical limits prevent the
line length from reaching to a saturated value, it is expected that the
length will be saturated eventually though. The line length for 30 mm/s
(not plotted in Fig.~\ref{vorlen}) decreases to zero in a period of 11
ms after stopping the injection of vortex rings.


In summary, we report experimental and numerical investigations of the
responses of a vortex--free oscillating obstacle sprinkled with vortex
rings. In the experiment, slow filling of superfluid $^4$He and
filtering of vortices enable us to produce a vibrating wire that is
effectively free of remanent vortices, and which cannot generate
turbulence. Using this vibrating wire, we successfully observed the
transition to turbulence triggered by free vortex rings. The numerical
simulations of vortex dynamics demonstrate that vortex rings can attach
to the surface of an oscillating obstacle and expand due to fast
boundary flow to form turbulence. Thus the combination of experimental
and numerical investigations reveals new insights into the phenomenon of
quantum turbulence.

\begin{acknowledgments}
The research was supported by a Grant-in-Aid for Scientific Research on
Priority Areas (Grant No.~17071008) from The Ministry of Education,
Culture, Sports, Science and Technology of Japan, and from JSPS
(Grant No.~18340109).
\end{acknowledgments}


\begin{thebibliography}{20}
\bibitem{ring_size} In the experiment, we observed a delay between the
		generation of vortex rings by a generator to the
		turbulent transition of a detector. We estimated the
		typical size of vortex rings to be about 0.7 $\mu$m from
		this delay time.
\end{thebibliography}

\end{document}